\documentclass[prb,twocolumn,floatfix,showpacs]{revtex4}
\usepackage{amsfonts}
\usepackage{amssymb,amsmath}
\usepackage{bm}
\usepackage{graphicx}
\begin{document}
\title{Incommensurate DDW order }
\author{Ivailo Dimov and Chetan Nayak}
\affiliation{Department of Physics and Astronomy, University of California at Los Angeles,
Los Angeles, California 90095-1547, USA}
\date{\today }
\begin{abstract}
We consider various incommensurate (IC) order parameters for electrons
on a square lattice which reduce to
$d_{x^2-y^2}$-density wave (DDW) order when the
ordering wavevector ${\bf Q}\rightarrow (\pi,\pi)$.
We describe the associated charge and current distributions
and their experimental signatures. Such orders can arise at the
mean-field level in extended Hubbard models. We compare the phase diagrams of
these models with experiments in the underdoped cuprates, where (1) DDW order
is a possible explanation of the pseudogap, and (2) there are experimental
indications of incommensurability.
We find various types of IC DDW and discuss their
possible relevance to the physics of the cuprates. Our main finding is that
IC DDW order is generally accompanied by superconducting order,
but the magnitude of the IC wavevector can be small.
A comparison with the analogous AF-ICSDW transition
is given.
\end{abstract}

\maketitle

\section{Introduction}\label{intro}

When a commensurate one-dimensional charge density wave (CDW) is doped,
the resulting charges may be viewed as defects in the CDW.
If the interactions between them are sufficiently strong
compared to their effective kinetic energy, then they will form an ordered lattice
(which can be stabilized by a crossover to 3D order if there are
many $1D$ chains coupled together)
of defects and the CDW will become incommensurate with the lattice.
This can be more energetically favorable than simply doping holes into the rigid band
formed in the presence of CDW order at a fixed wavevector. The reason is that,
in the latter case, we must pay the single-particle energy gap of the CDW.
In a two-dimensional $d_{x^2-y^2}$-density wave (DDW)
state, however, there are nodes in the order parameter, so it is relatively painless
for doped holes to simply go to the nodes, which expand into Fermi pockets
(in quasi-$1D$ systems, however, this might not be possible, so doping must
lead to incommensurability \cite{Schollwock03}).
However, this cannot continue indefinitely if the DDW state is stabilized by
approximate nesting, since the Fermi surface (FS) would eventually move away
from the nesting wavevector. Thus, incommensurate DDW order
is a strong possibility, at least over some range of hole dopings. In this paper,
we explore this possibility.

Near half-filling, the Hubbard model, the $t-J$ model, and generalizations
of these appear to have many phases which are nearby in
energy \cite{Schollwock03,numerics}.
Thus, small perturbations can strongly influence the competition between them.
Consequently, it should not be too surprising that
experiments on the cuprate superconductors have also uncovered evidence
for a cornucopia of phases, particularly on the underdoped side of the phase diagram, which
appear in particular materials and for certain values of the doping level, temperature, magnetic
field, etc. \cite{expts}.
Depending on the interfacial energies between these phases,
one way in which their competition can be resolved
is through the formation of stripes \cite{stripes-theory}
or other inhomogeneous patterns of microscale phase-separation.
Neutron scattering and scanning tunneling microscopy experiments point towards this possibility\cite{stripes-expts,stm}.
Furthermore, as the cuprates are doped, their Fermi surfaces
evolve away from the commensurate nesting wavevector $(\pi,\pi)$,
so we would expect translational symmetry-breaking order
parameters to occur at incommensurate wavevectors.
Hence, it would seem
important for any theory of the pseudogap to
incorporate the tendency towards incommenuration,
which we take as further impetus to study
incommensurate (IC) order parameters
related to DDW order.
 
In this paper we consider various ways in which $d_{x^2-y^2}$-density wave
(DDW) order can go incommensurate. Before turning to this
discussion, we review the reasons for expecting
DDW order to occur in the pseudogap regime of the cuprates.
It has been suggested\cite{ddw,Nayak_lddw}
that competition between DDW order and 
$d$-wave superconductivity (DSC) can explain many of the anomalous properties of the pseudogap\cite{TimuskStatt}.
The DDW order parameter is given by
\begin{equation}
\label{cop}
\langle{c}^{\dagger}_{{\bf k}\sigma}c^{}_{{\bf k+Q}\sigma}\rangle
= i\Phi{f}({\bf k})
\end{equation}
with ${f}({\bf k})=\cos(k_x)-\cos(k_y)$ and a {\it commensurate}
(C) ordering wavevector ${\bf Q}=(\pi,\pi)$.
The particle-hole condensate in \eqref{cop} breaks translation
by one lattice spacing, parity, and time-reversal
but is invariant under the combination of any two of the above symmetries.
In real space it is represented by currents
which alternate from one plaquette to the next.

One motivation for considering DDW order is to
explain \cite{Sumanta} the abrupt depletion of the superfluid density
below the critical $x_c\approx 0.19$ in Bi2212 and similar dopings in other materials \cite{Loram,superfluid}.
Because the order has a $d$-wave momentum dependence,
its onset would explain ARPES measurements
above $T_c$ in the pseudogap \cite{ddw-arpes}.
Furthermore, since DDW is a spin-singlet condensate,
it would also explain the suppression of the spin susceptibility in NMR measurements and
opening of a spin gap in inelastic neutron scattering \cite{ddw}.
Also, the depletion of the c-axis conductivity \cite{c-axis},
whose contribution mostly comes from the antinodal
regions, would also be naturally explained by the development of DDW order,
which gaps these regions. Direct attempts to measure DDW order through
neutron scattering have neither ruled it out \cite{Stock} nor unambiguously verified
its presence, but there are intriguing suggestions that it may be present\cite{Mook}.
Thus, it is natural to ask whether DDW order is compatible with
incommensurability. (For a  discussion of the experimental signatures of DDW
order, the reader is referred to the literature
\cite{ddw,Sumanta,ddw-neutron,ddw-arpes,ddw-Hall,Hall-Boebinger,ddw-IR-Hall,IR-Hall-Drew}. The case of DDW in $La_{2-x}Ba_xCuO_4$ is discussed in\cite{ddw-SO-lsco}. For numerical evidence of the existence of DDW in other systems, one is referred to\cite{ddw-num}.)

The plan of our paper is as follows. In section \ref{susc} we develop
some preliminary intuition about the energetics of
IC density wave order by analyzing the DDW susceptibility of a
Fermi liquid (both the nested and non-nested cases).
In section \ref{c-ic} we review some of the theory of
commensurate (C) DDW order and show various ways to make it incommensurate.
We also suggest the kind of microscopic physics that
could lead to C and IC DDW, and in addition discuss possible experimental signatures of IC DDW. Finally, in section \ref{num}, we compute the mean-field phase diagrams of an extended Hubbard
model. Our main results are that (a) IC DDW develops for a wide range of parameters, but
(b) the IC wavevectors can be very small. Their relevance to the cuprates
may, at most, be for a rather narrow set of dopings.
We also offer a physical explanation of our results, which sheds
some light on the nature of C-IC DDW transition in extended Hubbard models.
We note that similar incommensurate order has been found in refs. \cite{Cappelluti99, Kee02}.

\section{Susceptibility Near Half-Filling}
\label{susc}

The easiest way to see that IC DDW  phases could arise in effective models
(which, in this paper, will, in turn, emerge
from microscopic extended Hubbard models with
correlated hopping terms\cite{NayakPiv};
see section \ref{micro} for details) is to consider the bare susceptibility
as a function of the incommensurability wavevector ${\bf q}$, where
${\bf Q} = (\pi,\pi)+\bf{q}$:
\begin{equation}
\label{suscep}
{\chi_0}({\bf q}) = \frac{1}{2\pi^2}\int_{\bf k}\left(f({\bf k})\right)^2
\frac{n_F(\epsilon_{\bf k+Q}-\mu)-n_F(\epsilon_{\bf k}-\mu)}{\epsilon_{\bf k}-\epsilon_{\bf k+Q}}.
\end{equation}
where $f({\bf k})$ is defined after eq. \ref{cop}.
Here, for simplicity, DDW is considered without the presence of any other order.
In eq. \ref{suscep}, $n_F$ is the Fermi occupation number, and the energy
dispersion contains nearest, $t$, and next-nearest neighbor, $t'$, hopping parameters,
$\epsilon_{\bf k}=-2t(\cos{k_x}+\cos{k_y})-4t'\cos{k_x}\cos{k_y}$. 
The DDW equivalent of the Stoner criterion is satisfied when the right-hand-side of this equation
approaches the inverse of the pertinent coupling $1/g$. This observation allows us to
interpret the positions of the peaks of ${\chi_0}({\bf q})$ as the wavevectors at which DDW order
is likely to occur as we lower the temperature. A similar analysis was done by Schulz\cite{Schulz} for the
case of AF and IC SDW order in the Hubbard model close to half-filling,
where perfect nesting at half-filling was asumed. In fact,
\begin{figure}[htb]
\centerline{\includegraphics[height=3.4in]{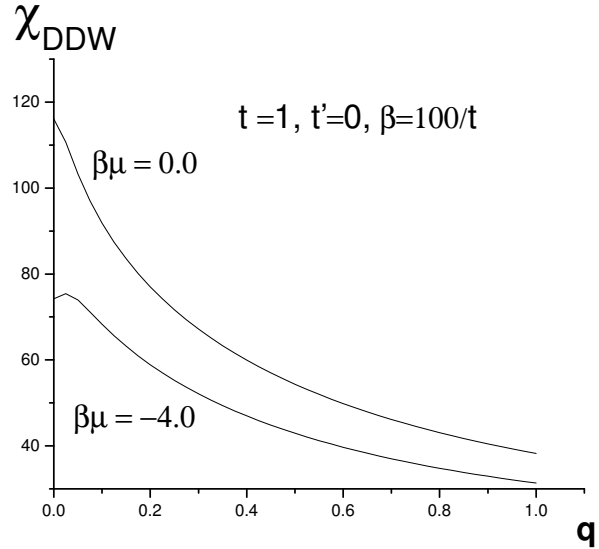}}
%\vskip 0.5cm
\caption{Susceptibility to DDW order. Away from half-filling, the peak of
${\chi^{}_0}({\bf q})$ is at nonzero ${\bf q}$. Here, $q$ is measured
along the zone diagonal in units of $(\pi/a,\pi/a)$, where $a$
is the lattice spacing.}
\label{suscfig}
\end{figure}
his results can be recovered by letting $f({\bf k})\rightarrow{1}$ in \eqref{suscep} if we set  $t' = 0$.

The support of $\left(f({\bf k})\right)^2$ in the
integrand in \eqref{suscep} is at $(0,\pi)$ and
symmetry-related antinodal points of the FS since this is where
the DDW order parameter is large. However, since these points are also the location of
van Hove singularities, the regions around them also dominate the AF
susceptibility. Thus, at least in the $t' = 0$ case, one expects that whenever
an IC SDW instability occurs, an IC instability
for DDW should also occur. Indeed, for both orders, the peak of ${\chi_0}({\bf q})$
occurs at ${\bf q}\neq{0}$ for low enough temperatures and sufficiently away from half-filling, e.g. if $|\mu\beta|\geq\mathcal{O}(1)$, as shown in Fig. \ref{suscfig}. 
Note that the same analysis repeated with any other density wave or spin-density wave order would yield an IC instability as long as the dominant channels of the modulation coincide with
Fermi surface points which give large contributions to the susceptibility. 
For instance, a $p_x$-density wave\cite{Nayak_lddw}
with wavevector ${\bf Q}=(0,\pi)+{\bf q}$
has a peak in ${\chi_0}$ for ${\bf q}\neq0$ away from half filling at $T\simeq0$, in the case $t' = 0$.

A natural question to ask is: to what extent does the IC DDW instability mimic the IC SDW one? To answer this, one needs to consider specific models as we do in \ref{micro}. However, two differences can be anticipated even beforehand. First, the DDW order parameter has nodes, while the AF order parameter does not.
%Therefore, while at mean-field level of the weak coupling Hubbard model nesting is necessary for the %onset of AF, the DDW order naturally arises even if $t' \neq 0$ in models of the Hubbard %type\cite{NayakPiv}.
Second,  in a more realistic model, one must consider competition with DSC, which also could change features of the transition to IC state. A better understanding of IC DDW order parameter is necessary to better understand the the situation away from half-filling.

\section{C vs. IC Order}\label{c-ic}

\subsection{C Order parameter}

Given that the susceptibility \eqref{suscep} points towards an IC instability,
the next issue is what IC DDW order would look like. Before addressing this
question, let us first summarize some features of the C-case.\cite{Nayak_lddw}
In the C case, a general singlet particle-hole condensate can be written as
(neglecting the ${\bf k}={\bf k'}$ contribution):
\begin{equation}\label{corder_k}
\langle{c}^{\alpha'\dagger}_{\bf k'}c^{}_{\alpha,{\bf k}}\rangle
=i\Phi{f}({\bf k})\,\delta^{\alpha'}_{\alpha}
\delta_{{\bf k'},{\bf k+Q}}.
\end{equation}
Here, $\Phi$ is the magnitude of the order parameter.
The wavevector $2{\bf Q}$ is a reciprocal lattice vector. The spin index $\alpha$
included above will be omitted in the future for simplicity unless clarity requires otherwise.
In  \eqref{corder_k}, the form factor $f({\bf k})$, which can transform non-trivially under
the point group of the 2-D lattice, is related to the angular momentum of the particle-hole
condensate. For DDW order, $f({\bf k})=\cos{k_x}-\cos{k_y}$ which corresponds to $l=2$ and
$d_{x^2-y^2}$ symmetry.  A $p_x$-density-wave with ${\bf Q}=(0,\pi)$ has $f({\bf k})=\sin{k_x}$,
which corresponds to $l=1$.

In real space, \eqref{corder_k} becomes
\begin{equation}\label{corder_x}
\langle{c}^{\dagger}_{\bf r'}c_{\bf r}\rangle
= \mathcal{V}({\bf r},{\bf r'})\,\Phi \cos({\bf Q\cdot{r}}) 
\end{equation}
where the vertices $\mathcal{V}$ of the orders discussed above are given by
\begin{equation}\label{vertices}
\begin{split}
-i\mathcal{V}^{\rm DDW}({\bf r},{\bf r'})&=\delta_{\bf r',r+\hat{x}a}+\delta_{\bf r',r-\hat{x}a}-
	\delta_{\bf r',r+\hat{y}a}-\delta_{\bf r',r-\hat{y}a}\\
-i\mathcal{V}^{p_x\text{DW}}({\bf r},{\bf r'})&=\delta_{\bf r',r+\hat{x}a}-\delta_{\bf r',r-\hat{x}a}.
\end{split}
\end{equation}
Note that the factors of $i$ in $\mathcal{V}^{DDW}$ and $\mathcal{V}^{p_x\text{DW}}$ signify that the 
corresponding phases break time-reversal, e.g. current flows along the bonds. Positive/negative signs in 
\eqref{vertices} represent current going in/out of the vertex.  Translation invariance, by default, is broken by 
any density wave order since there is a preferred vector ${\bf Q}$, but in the C case, invariance
under a {\it subset} of the lattice group is retained. DDW order, for example, is invariant under translation by
linear combinations of even reciprocal lattice vectors along each bond direction.  The current along the
bonds is given by:
\begin{equation}
\label{current}
{\bf j}_{{\bf r}\leftrightarrow{\bf r}+\hat{s}a}
= it(\langle{c}^\dagger_{\bf r}c^{}_{{\bf r}+\hat{s}a}\rangle-
\langle{c}^\dagger_{{\bf r}+\hat{s}a}c^{}_{\bf r}\rangle)
\end{equation}
Here $\hat{s}$ is along either the $\hat{x}$- or $\hat{y}$-direction.
Because of the equivalence of ${\bf Q}$ and $-{\bf Q}$ in the C case, the current  is simply
${\bf j}_{{\bf r}\leftrightarrow{\bf r}+\hat{s}a}
\sim\cos({\bf Q}\cdot\hat{s}a)\sim\langle{c}^\dagger_{\bf r}c_{{\bf r}+\hat{s}a}\rangle$. In
the DDW case this results in an alternating plaquette current/checkerboard magnetic field
patterns.\cite{Nayak_lddw}

 \subsection{IC Order Parameter}\label{icop}
\begin{figure}[htb]
\centerline{\includegraphics[height=3.0in]{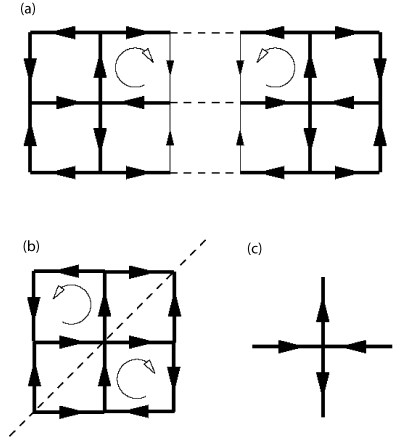}}
%\vskip 0.5cm
\caption{(a) Bond-oriented DDW domain wall. 
(b) Diagonal DDW domain wall.
(c) Current vertex of DDW.
There is no current on dashed lines lines. Arrowless currents are regions of
increased kinetic energy. Thick arrows represent one (arbitrary) unit of current. 
Thin lines represent half a unit of current.}
\label{fig:domains}
\end{figure}

To obtain an IC version of DDW order, one first notes that DDW is chiral and hence breaks the Ising
symmetry which reverses all the clockwise plaquette currents into counter-clockwise and vice versa.
Therefore one can construct anti-phase domain walls as in Fig.\ref{fig:domains}. A simple
caricature of a bond-oriented domain wall is given in Fig.\ref{fig:domains}a.
Its real space representation is
\begin{equation}\label{domain_single}
\langle{c}^{\dagger}_{\bf r'}c^{}_{\bf r}\rangle
=\mathcal{V}^{\rm DDW}({\bf r},{\bf r'})\Phi\cos{\bf Q\cdot{r}}\times(2\Theta(r_x)-1)
+\mathcal{V}^{\rm IC}
\end{equation}
where $\Theta(x)$ is the step function, so
\begin{equation}\label{step_FT}
2\Theta(x)-1=\sum_\text{n}\text{A}_{2\text{n}+1}\sin\!\left(\frac{(2\text{n}+1)\pi{x}}{L}\right)
\end{equation}
with $\text{A}_{2n+1}=4/\pi({2n+1})$. At the midpoint of the domain wall, the currents in the
$x$-direction vanish while the currents in the $y$-direction are halved.
The diagonal domain wall depicted in figure \ref{fig:domains}b
has a similar representation.  Note that because of the halving of the  $y$-direction currents at the midpoint of the bond-domain wall in \eqref{domain_single}, we need to include  the term:
\begin{equation}
\mathcal{V}^{\rm IC}
= \mathcal{V}^{\rm R}({\bf r},{\bf r'})\delta_{r_x,0}+\mathcal{V}^{\rm L}({\bf r},{\bf r'})\delta_{r_x,1}
\end{equation}
where the new vertices $\mathcal{V}^{\rm L, R}$ are given by:
\begin{equation}\label{vert_ic}
\begin{split}
-i\mathcal{V}^{\rm L} & = \delta_{\bf r',r-\hat{x}a} -
	\frac{1}{2}\delta_{\bf r',r+\hat{y}a}-\frac{1}{2}\delta_{\bf r',r-\hat{y}a}\\
-i\mathcal{V}^{\rm R} & = \delta_{\bf r',r+\hat{x}a} -
	\frac{1}{2}\delta_{\bf r',r+\hat{y}a}-\frac{1}{2}\delta_{\bf r',r-\hat{y}a}.
\end{split}
\end{equation}
Such vertices seem necessary for IC transitions of current density waves. One could imagine situations in which the transition to a disordered state is driven by thermal proliferation of such vertices, similar to earlier analysis of the 6-vertex model\cite{Sudip-6-vertex}. Although the analysis of such a transition is an interesting topic on its own, we will not pursue it in this paper.  We are more interested in the energetics of the IC phase, so a long-wavelength mean-field treatment is sufficient.

To analyze the C-IC transition, one then considers an array of alternating domain walls
with average separation $L$. Then, in the single harmonic approximation one retains only the
$\text{n}=0$ term in \eqref{step_FT} and identifies $q=\pi/L$ with the incommensurability wavevector. The new short-range vertex contributions \eqref{vert_ic} are also neglected in the lowest harmonic
approximation.
The neglect of higher harmonics should not change the transition temperature\cite{Kotani} or the
value of the critical doping for the onset of IC order since $\Phi$ is small and the corrections
are of higher order in $\Phi$. An incommensurate instability in a clean system occurs when $L$ becomes finite. (In a disordered system, it occurs when when $L<L_{d}$,
where $L_d$ is a characteristic length scale for the disorder.)

Within the single harmonic approximation, an array of anti-phase domain walls with separation 
$|{\bf q}|~=~|{\bf Q-(\pi,\pi)}|~=~\pi/L$, looks simpler in momentum space:
\begin{equation}\label{icorder_k}
\langle{c}^{\dagger}_{\bf k'}c_{{\bf k}}\rangle
=i\frac{\Phi}{2}{f}({\bf k})(\delta_{{\bf k'},{\bf k+Q}}+\delta_{{\bf k'},{\bf k-Q}}).
\end{equation}
Note that in the limit $|{\bf q}|\rightarrow0$, the above reduces to \eqref{corder_k}.
One could have guessed the result  \eqref{icorder_k} by writing 
$2\delta_{{\bf k'},{\bf k+Q}}=\delta_{{\bf k'},{\bf k+Q}}+\delta_{{\bf k'},{\bf k-Q}}$ in the commensurate case,
and then let ${\bf Q}$ go incommensurate. This suggests other ways to get IC order.
In particular the oddness of DDW order with respect to rotations by $\pi/2$ and
transpositions about, say, the $(\pi,\pi)$ direction lead to the following example of a DDW {\it checkerboard pattern}:
\begin{equation}\label{checkerboard}
\langle{c}^{\dagger}_{\bf k'}c_{{\bf k}}\rangle
=\frac{i}{4}\Phi{f}({\bf k})(\delta_{{\bf k'},{\bf k+Q}}+\delta_{{\bf k'},{\bf k-Q}}) 
-\{ {\bf Q}\rightarrow\mathcal{O}({\bf Q} )\}
\end{equation}
where the operation $\mathcal{O}({\bf Q})$ can be either a transposition, or a rotaion by $\pi/2$.
The checkerboard pattern in \eqref{checkerboard} corresponds to simply superimposing two
domain walls rotated by $\pi/2$ with respect to each other.

In a similar spirit, one could 
also investigate the possibility of {\it non-topological} domain walls\cite{KivDemleretc}, e.g. IC domains as the above, superimposed with an uniform order, so the overall order parameter does not change chirality. However, as we will see in Sec. \ref{num}, much of the physics of the C-IC transition will already be transparent from the simpler choices \eqref{icorder_k} and \eqref{checkerboard} of order parameter. 

Finally, note that within the single harmonic approximation, \eqref{icorder_k} and \eqref{checkerboard} are current-conserving only to lowest order in $q$. For the purpose of clarifying the energetics of the IC order close to the C-IC transition, or in the cases where the magnitude of the incommensurability is small, this lowest order approximation is good enough. Our numerical results, Sec. \ref{num}, {\it a posteriori} show that we are in such regime.

\subsection{Microcopic Models}\label{micro}

We will consider the above order parameters in the context of the following Hamiltonian
\cite{NayakPiv}:
\begin{multline}
\mathcal{H}=-t_{ij}\sum_{\left\langle i,j\right\rangle }\left(
c_{i\sigma }^{\dagger }c_{j\sigma }^{}+\text{h.c.} \right)
- {t_c} {\hskip -0.3 cm}
\sum_{\substack{ \left\langle i,j\right\rangle ,\left\langle i^{\prime},j\right\rangle \\i\neq i^{\prime} }}
c_{i\sigma }^{\dagger}c_{j\sigma }^{}c_{j\sigma }^{\dagger }c_{i^{\prime}\sigma }^{}\\
+ U\sum_in_{i\uparrow }^{}n_{j\downarrow}^{}
+V\sum_{\left\langle i,j\right\rangle }n_i^{}n_j^{}
\label{micro-Ham}
\end{multline}
In this formula, $t_{ij}$ is hopping with $t_{ij}=t$ for nearest
neighbors, $t_{ij}=t^{\prime }$ for next nearest neighbors and $t_{ij}=0$
otherwise. $t_c$ is a correlated hopping term which simultaneously hops an electron
from site $j$ to site $i$ and hops an electron from $i'$ into the
vacated site $j$.
The on-site and nearest-neighbor repulsions are, respectively,
$U$ and  $V$. The
indices $i,j$ signify lattice sites and $\sigma $ the spin.

We will consider this Hamiltonian at the mean-field level. At this level, the
energetics of DDW order is the same as for a `BCS-reduced' Hamiltonian for DDW order of
the form:
\begin{equation}\label{ddw-reduced}
\mathcal{H}_{DDW}=-g_\text{DDW}\int\limits_{\bf k,k'}f({\bf k})f({\bf k'})c^\dagger_{{\bf k+Q}\sigma}c_{{\bf k}\sigma}
c^\dagger_{{\bf k'}\sigma'}c_{{\bf k'+Q}\sigma'},
\end{equation}
with $g_\text{DDW}=24t_c+8V$. A similar reduced Hamiltonian for $d$-wave
superconductivity has $g_\text{DSC}=12t_c-8V$.
Note that the processes responsible for both DSC and DDW are essentially kinetic.

A Hamiltonian of the form (\ref{micro-Ham}) has one important virtue: it has a $d$-wave
superconducting ground state over a range of dopings (and  an antiferromagnetic ground
state at half-filling). Since this is the {\it sine qua non} for any description of the cuprates,
we believe that this is a good starting point. One may object that correlated hopping terms
have to be put in by hand in a microscopic Hamiltonian.
However, they naturally arise even from the one-band Hubbard model away from half-filling,
where their contribution can be computed in $t/U$ perturbation theory:
\begin{equation}
t_c \simeq x\frac{t^3}{U^2}
\end{equation}
At dopings $x=0.1$, for $t/U\sim0.2$, we would have
$g_\text{DDW}=2g_\text{DSC}\sim 0.1t=J/2$.
Thus, not far away from half-filling, correlated hopping terms are
certainly not negligible.

\subsection{Mean Field Theory}\label{mft}

Let us generalize the mean field description of C DDW to the IC case. The path we will take is similar to that in Sec. \ref{icop}, where we wrote down the {\it commensurate} order parameter without making use of the ${\bf Q}\rightarrow{\bf -Q}$ symmetry, and as a result the generalization to incommensurate ${\bf Q}$ was straightforward. We thus obtain a functional $\mathcal{F}(\mu, \Delta_i, {\bf q})$, equation \eqref{freen}, whose minimization yields self-consistently both the order parameters $\Delta_i$ and the deviation from commensurability ${\bf q} = {\bf Q} - (\pi,\pi)$. The only subtlety, as we discuss at the end of this section, is that although $\mathcal{F}$ reduces to the mean-field  free energy in the C phase, in the IC phase it is {\it not} the free energy even within the single harmonic approximation. Instead, minimization of $\mathcal{F}$ is equivalent to a variational search of the ground state.

Our derivations generalize the treatment of Nayak and Pivovarov \cite{NayakPiv}
of competition of C DDW with DSC and possibly AF. 
The C DDW mean field Hamiltonian we start with is:
\begin{equation}\label{ham_ddw}
\mathcal{H}_{\text{DDW}}=\int\limits_{\bf k\in{\text{RBZ}}}
\{ i\frac{W_{\bf k}}{2}c^{\dagger}_{{\bf k}\sigma}
c_{{\bf k+Q}\sigma}+
{\bf Q}\rightarrow{\bf -Q}\}
+\{\text{h.c.}\}
\end{equation}
where $W_{\bf k}=\frac{W_0}{2}f({\bf k})$.

 Note that in \eqref{ham_ddw} the  ${\bf Q}\rightarrow{\bf-Q}$ symmetry for {\it incommensurate} ${\bf Q}$ is not equivalent to the Hermiticity of the Hamiltonian due to the ${\bf k}$-dependance of $W_{\bf k}$ (e.g. $\langle{c}^\dagger_{\bf k}c_{\bf k+Q}\rangle = iW_{\bf k}\neq \langle{c}^\dagger_{\bf k}c_{\bf k-Q}\rangle^* = -iW_{\bf k - Q}$ for IC {\bf Q}). This is unlike the case of AF order, where the interaction is simply
\begin{equation}\label{ham_sdw}
\mathcal{H}_{\text{AF}}=\int\limits_{\bf k\in{\text{RBZ}}}\phi
(c^\dagger_{{\bf k}\uparrow}c_{{\bf k+Q}\uparrow}-c^\dagger_{{\bf k}\downarrow}c_{{\bf k+Q}\downarrow})+\text{h.c.}
\end{equation}
and we have $\langle{c}^\dagger_{{\bf k}\sigma}\tau^{\sigma\sigma'}_3c_{{\bf k+Q}\sigma'}\rangle = \phi = \langle{c}^\dagger_{{\bf k}\sigma}\tau^{\sigma\sigma'}_3c_{{\bf k-Q}\sigma'}\rangle^*$. 

In order to check with earlier results on the C-IC transition of AF order in the {\it weak coupling} Hubbard model\cite{Schulz} (see Sec. \ref{susc}), we will also include a term like \eqref{ham_sdw} in our full reduced Hamiltonian. Of course, in the strong coupling, $U>t$, limit, one is not justified in treating AF as a Fermi surface instability, especially close to half filling. A $t-J$ description is more appropriate in this case.  However, as mentioned in Sec. \ref{micro}, both DDW and DSC are essentially kinetic in origin within the extended Hubbard model, so even in the strong onsite repulsion limit, there is a good justification to use mean field theory to describe their competition away from half filling. 

In the C case, the standard DSC term in our reduced Hamiltonian is:
\begin{equation}\label{ham_dsc} 
\mathcal{H}_{\text{DSC}}=\int\limits_{\bf k\in{\text{RBZ}}}\Delta_kc^\dagger_{{\bf k}\uparrow}c^\dagger_{{\bf -k}\downarrow}+\text{h.c.}
\end{equation}
where $\Delta_{\bf k}=\frac{\Delta_0}{2}f({\bf k})$. Using a Nambu basis $\Psi_{\bf k}\equiv(c^\dagger_{{\bf k}\uparrow},
c^\dagger_{{\bf k+Q}\uparrow},c_{{\bf -k}\downarrow},c_{{\bf -k-Q}\downarrow})$, the full Hamiltonian  $\mathcal{H}=\mathcal{H}_{\text{kin}}+\mathcal{H}_{\text{AF}}+\mathcal{H}_{\text{DDW}}+\mathcal{H}_{\text{DSC}}$ 
can be rewritten as 
$\mathcal{H}-\mu{N}=\int_{\text{RBZ}}\Psi^\dagger_{\bf k}\cdot{\bf A}\cdot\Psi_{\bf k}$ where:
\begin{equation}\label{matrix}
{\bf A}\equiv
\begin{pmatrix}
\epsilon_{\bf k}-\mu & iG_{\bf k}+\phi & \Delta_{\bf k} & 0\\
c.c. & \epsilon_{\bf k+Q}-\mu & 0 & \Delta_{\bf -k-Q}\\
c.c. & 0 & -\epsilon_{\bf k}+\mu & -iG_{\bf -k-Q}+\phi\\
0 & c.c. & c.c.& -\epsilon_{\bf k+Q}+\mu
\end{pmatrix}
\end{equation}
with $G_{\bf k} = (W_{\bf k} - W_{\bf k+Q})/2$.
The action derived from the above Hamiltonian yields, upon integrating the Fermions out, the free energy in the ordered state:
\begin{equation}\label{freen}
\mathcal{F}_{\bf q} = \mathcal{F}_{\text{quad}}+\sum_{s=\pm1}\int\limits_{{\bf k}\in\text{RBZ}}
\{(s\epsilon_{\bf k} -\mu) -\frac{2}{\beta}\ln2\cosh(\frac{\beta E_s}{2}) \}
\end{equation}
with  ${\pm}E_s, s=\pm1$ the energy eigenvalues of \eqref{matrix}, and 
\begin{equation}\label{f_quad}
\mathcal{F}_{\text{quad}}=\frac{\phi^2}{g_{\text{SDW}}}+\frac{W^2_0}{g_{\text{DDW}}}+\frac{\Delta^2_0}{g_\text{DSC}}. 
\end{equation}

If we assume ${\bf Q}\rightarrow{\bf -Q}$ symmetry, $E_s$ would be given by:
\begin{equation}\label{eval_c}
E_s({\bf k}) =\sqrt{\Delta^2_{\bf k} + [\mu-\epsilon_+({\bf k})+
s\sqrt{\bar\Phi^2_{\bf k}+\epsilon^2_-({\bf k})}]^2}
\end{equation}
where $\epsilon_\pm({\bf k})\equiv(\epsilon_{\bf k}\pm\epsilon_{\bf k+Q})/2$
and $\bar\Phi^2_{\bf k}\equiv{W}^2_{\bf k}+\phi^2$.

However, in the IC case, ${\bf Q}\rightarrow{\bf -Q}$ need not hold, in which case the above would generalize to:
\begin{widetext} 
\begin{equation}\label{eval_ic}
\begin{split}
E_s({\bf k}) &=\{f^2_+({\bf k})\Delta^2_{\text{DSC}}+(\mu-\epsilon_+({\bf k}))^2+\bar\Phi^2_{\bf k}+\epsilon^2_-({\bf k})+
s\sqrt{4(\mu-\epsilon_+({\bf k}))^2(\bar\Phi^2_{\bf k}+\epsilon^2_-({\bf k})) +
D_{\bf q}(\Delta^2_{\text{DSC}},\bar\Phi^2_{\bf k},\epsilon_{\bf k},\mu)
}
\}^\frac{1}{2},\\
D_{\bf q}&\equiv[4\tilde{f}_{\bf k}\bar\Phi^2_{\bf k}+f^2_-({\bf k})
\{ (\epsilon_{\bf k}-\mu)^2-(\epsilon_{\bf k+Q}-\mu)^2\}]\Delta^2_\text{DSC}+
f^4_-({\bf k})\Delta^4_\text{DSC}.
\end{split}
\end{equation}
\end{widetext}

The effect of a nonzero incommensurability ${\bf q}$ can be seen not only in the appearance of the terms
$f^2_\pm({\bf k})\equiv(f^2({\bf k})\pm{f}^2({\bf k+Q}))/2$ and  $\tilde{f}_{\bf k}\equiv(f({\bf k})+{f}({\bf k+Q}))/2$, but also in the DSC-and doping-dependent
function $D_{\bf q}$. 

Our final step before letting ${\bf Q}$ be incommensurate is to pass the integration in \eqref{freen} from RBZ to BZ:
\begin{equation}\label{rbz_bz}
\mathcal{F} - \mathcal{F}_{\rm quad} = \int\limits_{RBZ}f_s(\mu,\beta, \Delta_i) = \frac{1}{2}\int\limits_{BZ}f_s(\mu, \beta, \Delta_i) 
\end{equation}
Once we convert the RBZ integrals to BZ integrals as in \eqref{rbz_bz}, we numerically minimize (see Sec. \ref{num}) the resulting `free energy' with respect to both the order parameters $\Delta_i$ and the deviation from commensurability ${\bf q}$.

\begin{figure}[htb]
\centerline{\includegraphics[height=3.0in]{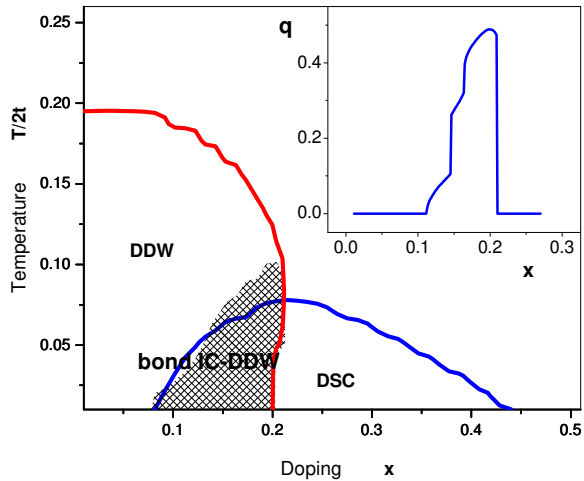}}
%\vskip 0.5cm
\caption{Phase diagram of DDW competing with DSC. Both $g_{\rm DDW}=0.05$eV and $g_{\rm DSC}=0.03$eV are large (compare to Figs. \ref{fig:ddwpd3} and \ref{fig:ddwpd4} ) $t=0.5$eV.
The next nearest neighbor hopping, $t'=-0.05$eV, is small compared to typical values fitted to ARPES\cite{arpes}, but IC order was nevertheless robust away from half-filling. Inset: IC wavevector $q$ at $T = 0.01$eV as a function of doping. Wavevectors are in
units of $\pi/a$, where $a$ is the lattice spacing.} %Note, however, that the difference in free energy of the different IC sectors was small, $\delta\mathcal{F}\sim0.1g_\text{DDW}$.}
\label{fig:ddwpd1}
\end{figure}

It must be emphasized that the  expression to be minimized in Sec. \ref{num} is {\it not} the true free energy of the system if ${\bf Q}$ is incommensurate. One way to see this is to note that the eigenvalues \eqref{eval_ic} are not invariant under ${\bf k}\rightarrow{\bf -k}$ if ${\bf Q}\neq(\pi,\pi)$ and therefore they cannot be interpreted as band energies. In a C-IC transition, the excitations split into a generally uncountably many bands even within the single harmonic approximation\cite{Kotani}. Therefore, the problem of finding the mean-field free energy is notoriously difficult even in the single harmonic limit. The calculation we are doing, on the other hand, is essentially a mean-field variational search for the ground state, and not for the excitations. A similar calculation was done by Schulz\cite{Schulz} for the case of the C-IC transition of the weak-coupling Hubbard model close to half-filling. By continuity of the true free energy\cite{Kotani}, one expects that the resulting variational estimates for  $q = |{\bf q}|$ would be accurate only close to the C-IC transition, where $q$ is small.

\section{Numerical Results}\label{num}
\begin{figure}[htb]
\centerline{\includegraphics[height=3.0in]{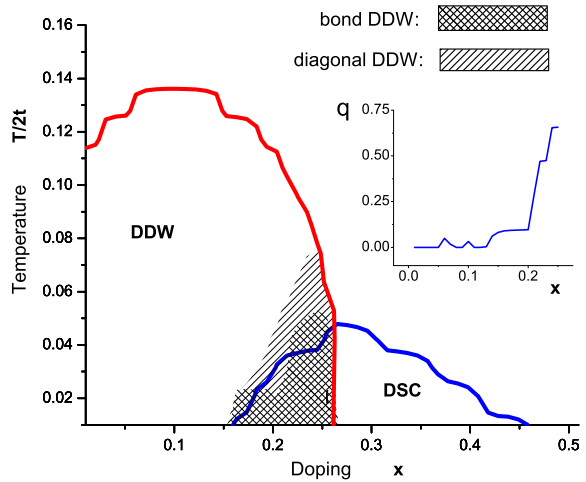}}
%\vskip 0.5cm
\caption{Again  $g_\text{DDW} = 0.04$eV, $g_\text{DSC} = 0.02$eV are `large' (note temperature scale) similar to Fig.  \ref{fig:ddwpd1}, with $t = 0.5$eV. However this time we taka a more realistic  $t' = -0.12$eV.
Again IC order is robust at dopings relevant for the pseudogap.  Inset: IC wavevector $ q$ at $T = 0.01$eV as a function of doping. See Fig. \ref{fig:qdw4-2tp012} for a more detailed view of $ q$ (and also Fig. \ref{fig:qdw4-2tp006} for the case of $t' = -0.06$eV ).}
\label{fig:ddwpd2}
\end{figure}

\begin{figure}[htb]
\centerline{\includegraphics[height=3.0in]{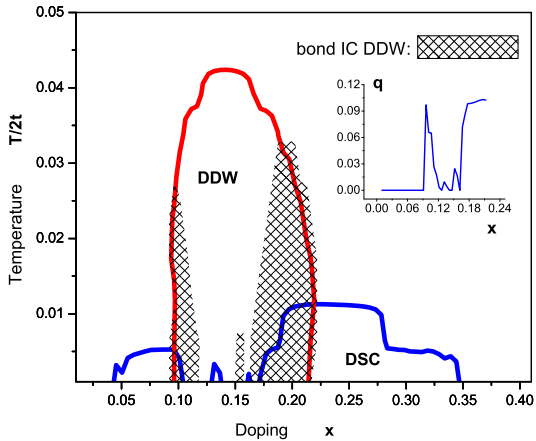}}
%\vskip 0.5cm
\caption{ $t=0.5$eV, $g_\text{DDW}=0.025$eV,
$g_\text{DSC}=0.01$eV. The next nearest neighbor hopping, $t'=-0.12$eV. Note the appearance of DSC proximate to `magic' dopings. Inset: the IC wavevector $q(x)$ at $T = 0.001$eV. See Fig.\ref{fig:qdw25-1tp012} for  comparison between $q(x)$ and $\mu(x)$. }
\label{fig:ddwpd3}
\end{figure}

To obtain the phase diagrams for the models in \eqref{freen}, for various fixed dopings $x$ and temperatures $T$, we minimize $\mathcal{F}=\mathcal{F}(\Delta_i, q)$ with respect to both the order parameters $\Delta_i$ (as was done previously in the C-case\cite{NayakPiv}) and the IC wavevector magnitude $q$. Each choice of IC order parameter (e.g. bond or diagonal) corresponds to a separate minimization procedure. To distinguish which IC order dominates at a particular point in $(x,T)$ space, we compared the free energies of the orders considered, and chose the IC phase of minimal energy. Note that,  as given in \eqref{freen}, the free energy is a function of  $\mu$, so to obtain the dependance on the doping $x$, we first perform a Legendre transformation and minimize instead $\mathcal{F}(x)=\mathcal{F}(\mu(x))+\mu(x)N$  where
the chemical potential is obtained from $1-x=\partial{f}/\partial{\mu}$ with
$f=\mathcal{F}/N_0$.

Our approach is slightly different than that previously used in the C-case\cite{NayakPiv}.  There, the competition between AF, DDW and DSC was analyzed, and some sensitivity on the details of the microscopics was found.
Phase diagrams resembling that of the cuprates were obtained.
The general trend was that any phase diagram
which included $d$-wave superconductivity also had DDW order competing
with it. By changing the bare coupling constants,
rather different phase diagrams could be obtained and
these generally did not have superconductivity.

In our treatment here, we do not consider the competition with AF
order, which is not a Fermi surface instability anyway
in the context of the cuprate superconductors.
In \eqref{freen}, we still include AF, but only in order to check with previous results\cite{Schulz} of IC AF order away from half filling in the {\it weak coupling}
$g_{AF} < t$ limit. On the other hand, the AF order appearing at half filling in the cuprates is a strong coupling phenomenon. A reduced Hamiltonian \eqref{ham_sdw} is the wrong starting point at half filling in the limit $U>t$. 

Therefore, we only consider competition between DDW and DSC and look for IC order in this case. All the phase diagrams constructed this way are robust in the sense that DDW and DSC are found to coexist at mean field level for almost all reasonable values of the coupling constants.

\begin{figure}[htb]
\centerline{\includegraphics[height=3.0in]{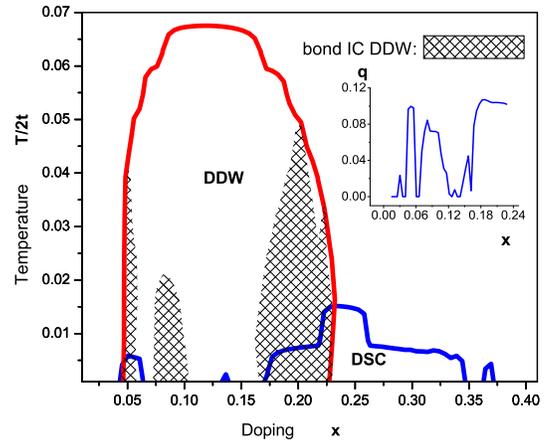}}
%\vskip 0.5cm
\caption{Same as Fig. \ref{fig:ddwpd3} except $g_\text{DDW}=0.03$eV. There is a distinct IC response (inset) at $x = 1/8$ and $x = 1/16$. See Fig. \ref{fig:qdw2-1tp012} for comparison between $\mu(x)$ and $q(x)$.}
\label{fig:ddwpd4}
\end{figure}

Sample phase diagrams are shown in Figs. \ref{fig:ddwpd1}-\ref{fig:ddwpd4}. In all of them, IC order is present at dopings $0.15 \lesssim x \lesssim 0.2$ where DDW and DSC compete. In the first two of these figures, we vary $t'$ at relatively high $g_{\rm DDW}$ and $g_{\rm DSC}$, while in the last two we explore the phase diagram for more realistic DDW and DSC couplings (compare the temperature scales to those in the cuprates). Note the complexity of Figs. \ref{fig:ddwpd3}, Figs. \ref{fig:ddwpd4}
compared to Figs. \ref{fig:ddwpd1},
Figs. \ref{fig:ddwpd2}. At realistic couplings,
the IC wavevector profile is complicated,
and furthermore DSC order appears in more than one doping region within the DDW dome. In fact, the appearance of incommensurability seems correlated with the appearance of DSC. A more systematic analysis to be discussed below reveals that generally the types of C-IC transitions that were produced in our phase diagrams fall into two categories: (a) those which correspond to motion of the bare Fermi surface away from commensurate nesting (shaded regions); 
\begin{figure}[htb]
\centerline{\includegraphics[height=3.0in]{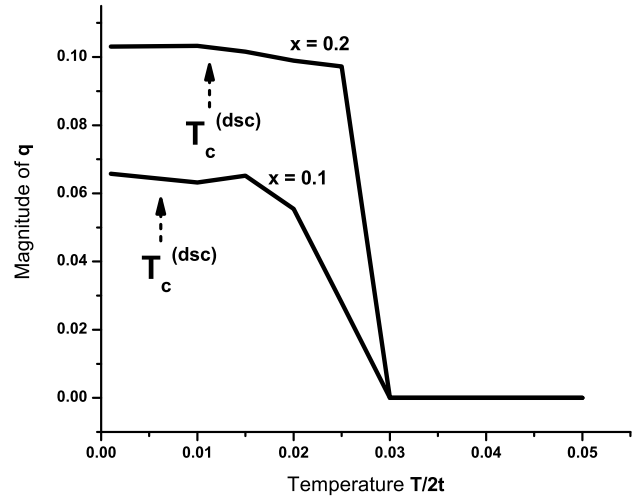}}
%\vskip 0.5cm
\caption{Dependance of incommensurability on temperature for regions in the phase diagram
in Fig. \ref{fig:ddwpd3} where DDW and DSC compete. Note that nonzero $q$ is found above as well as below the DSC critical temperature.}
\label{fig:qofT}
\end{figure}
(b) those which appear near certain `magic' filling fractions.  
Case (a) is similar to the familiar C-IC mechanism
discussed by Schulz\cite{Schulz}, while case (b) appears to be a consequence of negative effective stiffness at special fractions, the exact nature of which at present is not clear. In both cases $q$ is independent of the onset of DSC, Fig. \ref{fig:qofT}, but because incommensurability relieves the DDW-DSC competition, the onset of DSC correlates to the presence of IC DDW.

\begin{figure}[htb]
\centerline{\includegraphics[height=3.0in]{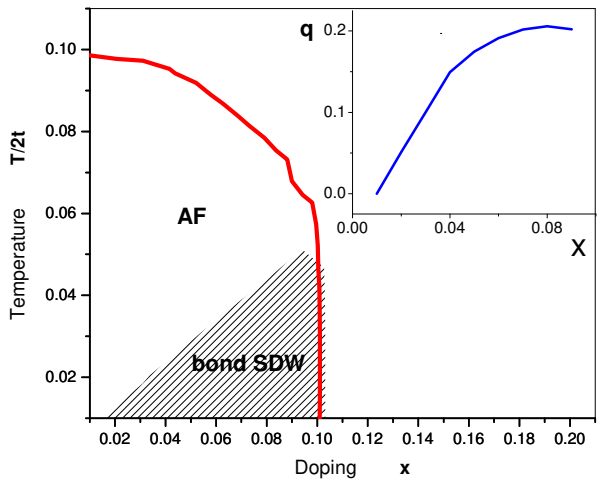}}
%\vskip 0.5cm
\caption{AF at perfect nesting, $t' = 0$. The results of Schulz\cite{Schulz} were recovered.  Here, we used $g_\text{AF}=2U=0.05$eV with $t=0.5$eV. Bond-oriented IC order was found to be energetically favorable, as a direct computation\cite{Schulz} in the weak repulsion regime would yield. {\it Inset}: As expected, the IC wavevector at $T=0$ is linear in $x$. }
\label{fig:afpd1}
\end{figure}

Let us first discuss (a). It is well known\cite{Schulz} that in the weak-coupling Hubbard model at $t' = 0$, the AF instability at half-filling is driven IC for any non-zero doping (see Sec. \ref{suscep}). The source of the C-IC transition can be traced to the approximate nesting of the bare Fermi surface (FS) at IC wavevectors away from half-filling.  Close to half-filling, this picture would produce a linear deviation from commensurability, $q\sim|\mu-\mu_c|\sim{x-x_c}$, Fig. \ref{fig:afpd1}. The incommensurability
(deviation from $(\pi,\pi)$) is horizontal or vertical,
not diagonal, as we would expect because
the bare susceptibility is more divergent due to a greater overlap between nested portions of the Fermi surface\cite{Schulz}.

A similar scenario happens in the case of DDW order.
However, two notable differences from the AF case are present. First, DDW is less sensitive to $t'/t$, since it is driven mainly by antinodal regions of high density of states (d.o.s.). Therefore, while there is no AF instability without nesting (at half-filling AF Fermi surface instability is only possible if $g_{AF}\gtrsim\mathcal{O}(t')$), DDW is still possible in this regime. This suggests that whenever the antinodal regions of the bare FS evolve away from the anti-nodal regons
(which happen to also be regions with a high density of states as a result
of the presence of van Hove singularities), a C-IC transition is likely to occur.

The second difference from the AF case is that DDW has nodes. This implies that the chemical potential in the ordered state need not be higher than the maximal DDW gap. As a result, DDW order can stay commensurate for wider ranges of doping. This is indeed what we find. For example, for all $t'/t \simeq 0$, we obtain $x^{\rm AF}_{c-ic} < x^{\rm DDW}_{c-ic}$ near zero temperature. In general, we find that the doping values $x_{c-ic}$ for which the above type of C-IC DDW is possible are unrealistically high in the context of the cuprates, if one considers more realistic bare FS, $|t'/t|\gtrsim0.25$.  Furthermore, in the IC DDW
case, there need not be
a simple relation $q\sim|\mu-\mu_c|\sim{x-x_c}$
between the doping and doping and deviation from
commensurability.

\begin{figure}[htb]
\centerline{\includegraphics[height=3.0in]{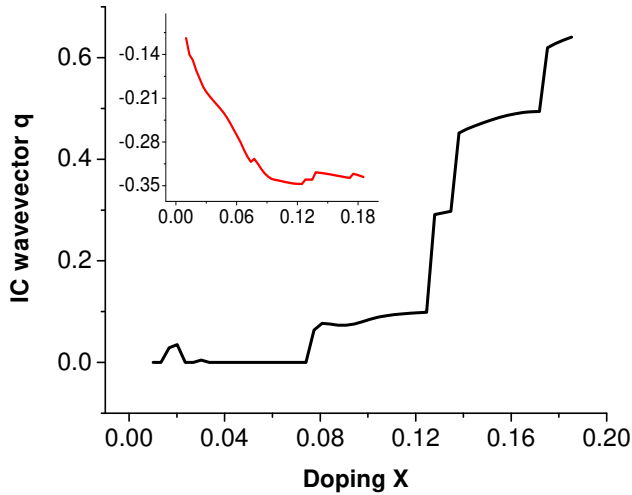}}
%\vskip 0.5cm
\caption{Comparison between $q(x)$ and $\mu(x)$ (inset) at $T = 0.01$ for $g_{\rm DDW} = 0.04$eV, $g_{\rm DSC} = 0.02$eV. Note that $|t'| = 0.06$eV which is smaller than the value taken in Fig. \ref{fig:ddwpd2}.}
\label{fig:qdw4-2tp006}
\end{figure}

\begin{figure}[htb]
\centerline{\includegraphics[height=3.0in]{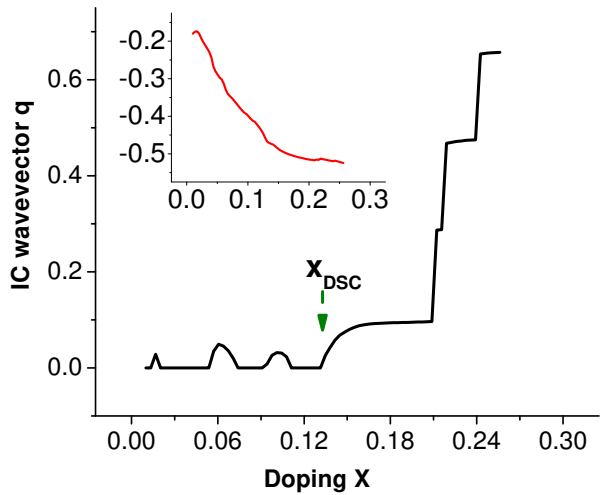}}
%\vskip 0.5cm
\caption{Same as Fig. \ref{fig:qdw4-2tp006} except this time $t' = -0.12$eV as in Fig. \ref{fig:ddwpd2}.}
\label{fig:qdw4-2tp012}
\end{figure}

\begin{figure}[htb]
\centerline{\includegraphics[height=3.0in]{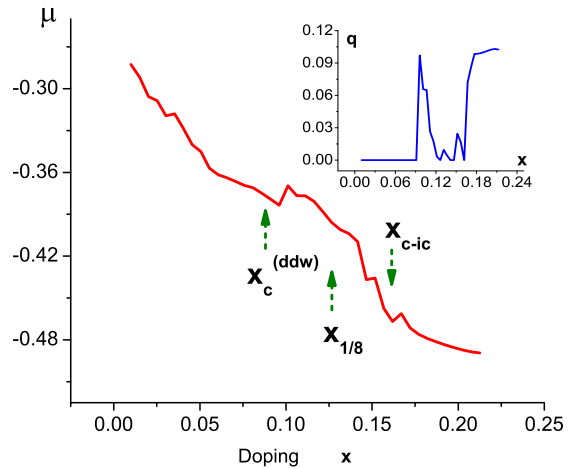}}
%\vskip 0.5cm
\caption{Comparison of $\mu(x)$ and $q(x)$ (inset) at $T = 0.001$eV corresponding to the parameters in Fig. \ref{fig:ddwpd3}. 
Although here the IC response is not at `magic' fractions, note that in Fig. \ref{fig:ddwpd3} there is DSC both at $x = 1/8$ and $x = 1/16$. }
\label{fig:qdw25-1tp012}
\end{figure}

\begin{figure}[htb]
\centerline{\includegraphics[height=3.0in]{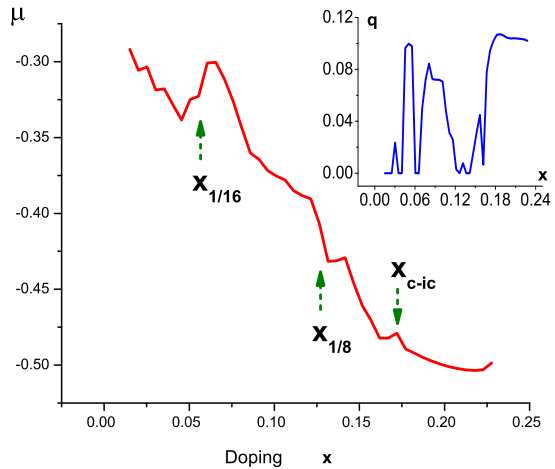}}
%\vskip 0.5cm
\caption{Same as Fig. \ref{fig:qdw25-1tp012} (and at the same temperature) except this time there is a clear C-IC transition at `magic' fractions (inset). At the same time $\mu(x)$ is non-monotonic at these dopings. The coupling constants  $g_{\rm DDW} = 0.03$eV and $g_{\rm DSC} = 0.01$eV are the same as in Fig. \ref{fig:ddwpd4}.}
\label{fig:qdw2-1tp012}
\end{figure}

\begin{figure}[htb]
\centerline{\includegraphics[height=3.0in]{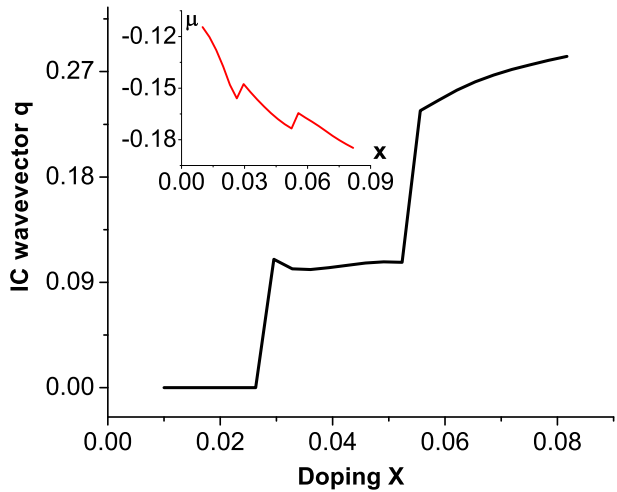}}
%\vskip 0.5cm
\caption{Example of the low temperature ($T = 0.001$ eV) staircase IC spectrum of AF around `magic' filling fractions. The parameters we have taken here are $g_{\rm AF} = -t' = 0.04$ eV, $t = 0.5$ eV. Inset: $\mu(x)$ is non-monotonic at both jumps of $q(x)$, but is smoothly decreasing whenever $q$ is pinned.}
\label{fig:qaf004T0tp004}
\end{figure}

Let us now discuss the C-IC transitions that are present in our phase diagrams proximate to `magic' filling fractions $x\sim1/2^n$. Although our understanding as to why they arise is incomplete, we would like to point out several interesting features about our results. First, to our knowledge, there is nothing special about the bare FS close to these filling fractions, likewise for the FS of the commensurate ordered state. The C-IC transition is not driven by nesting, since upon variation of $t'$ we see similar signatures proximate to the `magic' dopings. Second, the magnitude of the IC wavevector is generally small (corresponding to periodicity of more than $30$ lattice spacings) and shows no clear dependance on $n$, Figs. \ref{fig:qdw4-2tp006} -\ref{fig:qdw2-1tp012}.
Third, we checked that similar behavior is present 
deep into the AF ordered state\footnote{The AF only limit was taken by simply letting $g_{\rm DDW}, g_{\rm DSC}\rightarrow0$ in the total free energy \eqref{freen}.}, Fig. \ref{fig:qaf004T0tp004}.  Unlike the DDW case, however, the IC lattice spacing in the AF case depends on
$n$: $q$ follows an increasing staircase pattern at `magic' dopings.
Note that our results for AF order seem to contradict Schulz's results\cite{Schulz} which predict a linear dependence of $q(x)$. 

\begin{figure}[htb]
\centerline{\includegraphics[height=3.0in]{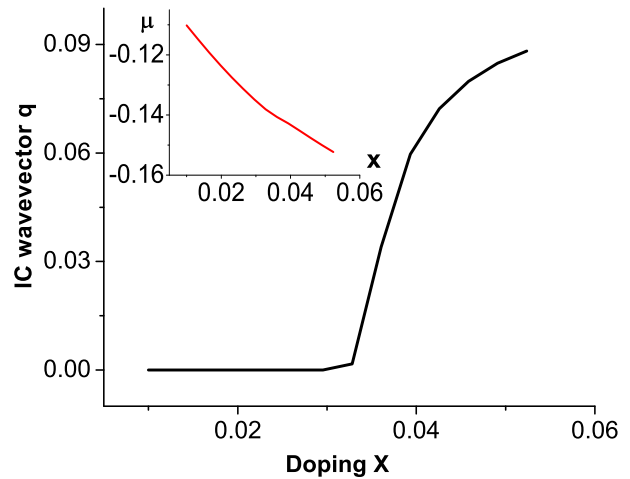}}
%\vskip 0.5cm
\caption{At higher temperatures $q(x)$ regains its linearity around $x_c$ as predicted by Schulz\cite{Schulz}. Here $T = 0.035$eV and all other parameters are as in Fig. \ref{fig:qaf004T0tp004}. Inset: $\mu(x)$ is monotonic throughout, suggesting that at higher temperatures, the C-IC transition is driven purely by IC nesting of the FS away from half-filling.}
\label{fig:qaf004T035tp004}
\end{figure}

The naive contradiction is easily resolved by realizing that Schulz's computation\cite{Schulz} of $q(x)$ relies on the Stoner criterion, which is strictly valid at $T_c$. Indeed, while deep into the ordered state the IC wavevector is stepwise increasing at `magic' fractions, at higher temperatures  we regain the linearity of $q(x-x_c)$,  Fig. \ref{fig:qaf004T035tp004}.  Moreover, the fact that both the bare and the gapped FS  in the C case are generally not nested at the the `magic' fractions suggests that the transition is not driven by a divergent susceptibility, but is first order. Indeed, at low temperatures, the chemical potential near the special dopings is non-monotonic for both the AF and the DDW+DSC diagrams. Similar behavior of $\mu(x)$ was found by Nayak and Pivovarov in the context of C DDW competing with DSC\cite{NayakPiv}.

A simple view of the energetics is given by the following Landau-Ginsburg free energy of two nodeless competing orders (the generalization to order parameters with nodes is straightforward):
\begin{multline}
\mathcal{F} = \int{d^2x} \Bigl(\rho_\psi |\nabla\psi|^2
+ \rho_\chi (\nabla\chi)^2 + K ({\nabla^2}\chi)^2\\
-m_\psi|\psi|^2 - m_\chi\chi^2 + \lambda|\psi|^2\chi^2
\Bigr)
\label{LGic}
\end{multline}
In this free energy, $\chi$ plays the role of the DDW order
parameter and $\psi$ the role of superconductivity.
When $\rho_\chi$ becomes negative, $\chi$ orders
at non-zero wavevector $q=\sqrt{-{\rho_\chi}/K}$.
This occurs at the dopings indicated in Figures
\ref{fig:ddwpd3} and \ref{fig:ddwpd4}.
When $\psi$ orders, this further favors incommensurability
if $\lambda>0$ since the competition between the two orders
is partially alleviated. This alleviation is a possible explanation of why
the onset of DSC at low $T$ is correlated to the presence of IC-DDW. 

The same phenomenon can be viewed from the opposite perspective: when DDW order
becomes incommensurate, there are regions in which it is suppressed (as in the domain walls
of Fig. 2). Superconductivity can occur more easily there. Hence, incommensurability
alleviates the competition between DSC and DDW order at the mean-field level, thereby enhancing
superconductivity. Of course, IC order appears to suppress superconductivity in the cuprates,
but this could be due to effects beyond a simple mean-field theory. 

When the superconducting order parameter becomes
strong enough, it may become energetically favorable
for phase separation to occur. We can guess when this
occurs from the usual Maxwell construction whenever the mean field chemical potential $\mu(x)$ does not vary monotonically.
This appears to occur at a few isolated `magic' doping values.
However, in a real system, proximity to a second order critical
point\cite{KivFradkinetal}, Coulomb repulsion\cite{KivReza},
or disorder could drive the system in a state of mesoscopic phase separation.
Moreover, the lattice pins any IC density wave order, so once the system is
IC in order to minimize the free energy, the deviation from commensurability $q$ would be pinned at a finite value.
Finally, it is important to remember that disorder could have
important effects. It would not only disorder the array
of domain walls, but could lead to
droplets of one phase in regions of another.

We stress that while the general trend towards IC order may be clear, many of the quantitative
aspects are not clear, in particular the precise copings at which DDW order becomes incommensurate.
it is not clear to us why such C-IC transitions happen near `magic' filling fractions.
The fact that similar behavior occurs in the case of AF order suggests that the tendency
of phase separation at `magic' filling fractions is a general trend for a variety of density wave order.
Because the AF only case was explored upon taking the $g_{DDW,DSC}\rightarrow0$
limit in the full free energy, it is not clear yet whether the tendency of phase separation
at `magic' filling fractions is purely a feature of density waves competing with other order parameters, or of the lattice. 
It might be worth reconsidering the AF case in the presence of other order parameters (such as DDW, DSC).
What makes our results surprising is that
unlike in the frustrated phase separation
picture for stripe formation, there is no Coulomb repulsion present.
Hence, there is no obvious connection between phase separation
and incommensurability. (Please note that while IC DDW
order can induce charge density wave order at
twice the wavevector, we have not
included this effect, which we expect to be small, in our calculation.)
Therefore, while Coulomb repulsion
may play a role in determining the CDW
spacing at $x = 1/8$ in $LaCuO_2$, in our calculation,
the phase separation at `magic' fractions can only arise as interplay between the lattice, competing order
physics and energetics. What exactly is the combination of the three deserves some future research\footnote{There is a neat way of thinking about IC DDW at finite $\mu$ as the analog of the Fulde-Ferrel-Larkin-Ovchinnikov (FFLO) state of DSC in a magnetic field parallel to the plane\cite{Kee02}. In the absence of next-nearest neighbor hopping, the two mean fields are related via a particle-hole transformation in the same way {\it commensurate} DDW is related to DSC at half-filling. However, while in the latter case one can rotate C DDW to DSC\cite{ddw}, in the case of finite $\mu$, the particle-hole transformation cannot interchange FFLO and IC DDW. While the mean-field Hamiltonians of IC DDW and the FFLO are related by a particle-hole transformation, the spectra are not -- in the IC DDW state there is a splitting to a generally uncountably many bands (if $Q/q$ is irrational as we discussed in Sec. \ref{num}) since the order parameter connects ${\bf k}$ and $\bf{k+Q}$ while in the case of DSC there is no such splitting. We are indebted to the authors of \cite{Kee02} for pointing out their work to us.}.

\section{Experimental Signatures}

According to the discussion following \eqref{LGic} it seems likely that within the DSC state a competing DDW order parameter is IC as it develops.  Hence, we briefly describe the experimental signatures that would distinguish between C and IC DDW order. Originally, C DDW was also named `hidden order' in the context of the pseudogap, because most of the experimental signatures associated with it are indirect. The magnetic field, created by the alternating currents around neighboring plaquettes, is too small to be detected by neutron scattering ($B\simeq1-30$ G for $\Delta_{\rm DDW}\sim30$ meV). At the same time, because of the symmetry of the C DDW magnetic field, there is no signature in NMR. The charge of the C DDW state is uniform so STM would not show signatures, unless they are associated with interference of nodal quasiparticles due to scattering off impurities. It has been argued recently\cite{ddw-quas-interf} that such interference disperses weakly enough to account for the pseudogap d.o.s. incommensurate modulations observed by STM. For a further discussion of the signatures of C DDW, the reader is referred to the literature
\cite{ddw,Sumanta,ddw-neutron,ddw-arpes,ddw-Hall,Hall-Boebinger,ddw-IR-Hall,IR-Hall-Drew}. 

An IC DDW is more easily detectable directly. The local electronic density of states would be modulated due to the IC order parameter, and hence could be observable by STM. Secondly, the IC modulations of the staggered currents would produce NMR line splitting of both the Cu and O atoms when the applied magnetic fields  are {\it perpendicular} to the a-b planes. This is in sharp contrast with IC SDW, where there is an NMR line-splitting due to spatially modulated local magnetization, which is directed {\it along} the a-b plane.

Let us estimate the modulated magnetic field strength arising from IC DDW  (the reader is referred to\cite{Colemanetal-orbAF} where similar considerations are applied to the case of orbital antiferromagnetism proposed to exist in $URu_2Si_2$). Suppose we have an array of domain walls, Fig. \ref{fig:domains}, along the y-axis. Deep between the domains the field is zero as in the case of C DDW.  Along a domain at $x = 0$, where the field is maximal, again most of the fields cancel except those created by  the vertical alternating currents of strength $I_0$ at $x = -a$ and those of strength $I_0/2$ at $x = a$. Equivalently this would be the field of a single vertical array of alternating currents of strength $I_0/2$ at $x = -a$:
\begin{equation}\label{Bmax}
 B_{\rm max} = \frac{\mu_0}{4\pi{x}_0} \sum\limits_{i}(-1)^i\frac{I_0}{2}(\cos(\theta_{i+1}) - \cos(\theta_i))
\end{equation} 
where $\cos{\theta_i} = [(i-1)a + y_0]/[\sqrt{((i-1)a + y_0)^2 + x^2_0}]$ and $(x_0,y_0) = (a, a/2),{\rm } (a, 0) $ for O, Cu atoms respectively. Clearly at $y_0 = 0$, Cu, atoms the field is zero by symmetry. For O atoms the sum can be performed yielding $4\pi{B_{\rm max}}/\mu_0 = 0.14 I_0$. In the dilute domain limit, this field is independent of the distance between the domains.

For a sinusoidal current modulation, $B_{\rm max}$ would depend on the modulation length $l_q = 1/q$. If $I(y,x) = I(y)sin(q x)$ then a good estimate of $B_{\rm max}$ is given by \eqref{Bmax} with $I_0$ replaced by $\delta{I}_0 = 2 q a I_0 = 2 I_0/n$ where $n$ is the periodicity of the modulation (in units of the lattice spacing). Therefore, for large $n$ we obtain $4\pi{B_{\rm max}}/\mu_0 = 0.24 I_0/n$.

\section{Conclusion}\label{concl}

Incommensurate order is an important possibility when competing orders are considered. At a
first-order transition the system is phase separated and hence necessarily inhomogeneous. If the competition between the orders is weak, coexistence is possible, but modulations of the competing orders are likely to be induced. The effects of the lattice, disorder, lomg-range Coulomb repulsion\cite{KivReza} and proximity to criticality\cite{KivFradkinetal} is to generally stabilize incommensurability on mesoscopic length scales. This is unlike the case of, say, the liquid to vapor phase transition where the competition results in phase separation on macroscopic length scales. 

In the case of DDW and DSC, the competition is certainly weak within the framework of the extended Hubbard models we have considered. Because the processes stabilizing both orders are of similar nature and are at the same time smaller than all of onsite repulsion, exchange and hopping, it does not cost much energy to convert DDW order into DSC and vice versa.
A simple resolution of the competition is for DDW order to become incommensurate.
We find that this occurs for some doping levels, but the incommensuration
tends to be relatively small.
Such IC DDW could, in principle, be observable at low temperatures by NMR since it creates an inhomogeneous magnetic field at O atoms {\it  perpendicular} to the ab-plane. Most of NMR experiments test for much stronger magnetic fields {\it parallel} to the ab-plane on the Cu atoms, and at higher temperatures such IC DDW phase would be smeared by thermal fluctuations, which
may exlpain why such IC order, if it exists, it hasn't been seen by experiment.

\acknowledgements

We would like to thank S. Brown, S. Chakravarty, R. Jamei, S. Kivelson, and E. Pivovarov for helpful discussions. Special thanks to E. Pivovarov for letting I. D. generalize some of his computer code. 
This work has been supported by the NSF under Grant No. DMR-0411800.

\end{document}